\documentstyle[12pt]{article}
\begin{document}
\newcommand{\siml}{\stackrel{<}{\sim}}
\newcommand{\simg}{\stackrel{>}{\sim}}
\baselineskip=1.333\baselineskip

\noindent
\begin{center}
{\large\bf A quantitative comparison between 
spike-train responses of Hodgkin-Huxley and
integrate-and-fire neurons}
\end{center}

\begin{center}
Hideo Hasegawa
\end{center}

\begin{center}
{\it Department of Physics, Tokyo Gakugei University  \\
Koganei, Tokyo 184-8501, Japan}
\end{center}
\begin{center}
{\rm (October 1, 1999)}
\end{center}
\thispagestyle{myheadings}
%
{\bf Abstract}   \par

Spike-train responses of single Hodgkin-Huxley (HH) 
and integrate-and-fire (IF) neurons 
with and without the refractory period, 
are calculated and compared.
The HH and IF neurons are assumed to receive
spike-train inputs 
with the constant interspike intervals (ISIs) 
and stochastic ISIs given by the Gamma distribution,
through excitatory and inhibitory synaptic couplings:
for both the couplings the HH neuron can fire
while the IF neuron can only for the excitatory one.
It is shown that the response to 
the constant-ISI inputs of the IF neuron
strongly depends on the refractory period and 
the synaptic strength
and that its response is rather different 
from that of the HH neuron.
The variability of HH and IF neurons depends not only
on the jitter of the stochastic inputs but also on
their mean and the synaptic strength.
Even for the excitatory inputs,
the type-I IF neuron may be a good substitute 
of the type-II HH neuron only in the limited 
parameter range

\vspace{0.5cm}


\vspace{1.5cm}
\noindent
{\it Corresponding Author}

\noindent  
Hideo Hasegawa \\ 
Department of Physics, Tokyo Gakugei University \\ 
4-1-1 Nukui-kita machi, Koganei, Tokyo 184-8501, Japan \\
Phone: 042-329-7482, Fax: 042-329-7491 \\
e-mail: hasegawa@u-gakugei.ac.jp

\baselineskip=0.75\baselineskip

\newpage
\noindent
{\bf I. Inroduction}

A neuron generates the action potential, which propagates
along the axon of a cell toward synapses exciting
neurons in the next stage.
There have been much theoretical studies on neurons which 
are responsible for encoding and decoding information carried
by action potentials \cite{Rieke96}.
In the most popular neuron models like 
the Hopfield model \cite{Hopfield82},
the output is described as a continuous variable 
which is slowly varying in time.  The output is
usually interpreted as short-time average of the rate
of action potentials. 
This is based on the experimental observation that
the mean firing rate depends on the applied stimulus
to motor and sensory neurons.

On the contrary, experimental evidences showing
that the detailed timing and organization of action 
potentials matter, have been reported
in many biological systems; sonar
processing of bats \cite{Suga83}, 
sound localization of owls \cite{Konishi92},
electrosensation in electric fish \cite{Carr86}, 
visual processing
of cats \cite{Eckhorn88} \cite{Gray89}, 
monkeys \cite{Rolls94} and human \cite{Thorpe96}.
These suggest the importance of studying how neurons 
make computations based on the action potential timing
with a resolution of the sub-millisecond range,
receiving and emitting spike trains.

Since Hodgkin and Huxley (HH)\cite{Hodgkin52} 
first proposed the reliable
model for squid giant axon, its property
has been intensively
investigated \cite{Nemoto75}-\cite{Matsumoto84}. 
The HH model has been widely adopted for 
an  investigation of biological systems with proper
modifications of conductance channels \cite{Arbib95}.
Because the HH model is described by the non-linear 
differential equations for four variables, its treatment
is not easy, and then various types of  simplified 
dynamical models have been proposed \cite{Albarbanel96}.
Among them, the simplest one is the integrate-and-fire (IF)
model, which has been employed for a study on many kinds
of subjects relevant to a single neuron 
as well as neural networks \cite{Maass97}.

It has frequently claimed that the IF model captures the
essentials of real 
neurons \cite{Hopfield95}-\cite{Troyer98}.  
On the other hand, some studies
have shown that the IF model is not 
realistic \cite{Feng97}-\cite{Brown99}.
Actually the IF and HH models show important differences
in their responses to applied, excitatory and inhibitory 
currents.
For an excitatory  (depolarizing) input dc current, $I_{\rm i}$,
the IF neuron, which is classified as the type I,
shows the self-excited oscillation with
an arbitrary low firing frequency, $f_{\rm o}$ \cite{Hodgkin48}. 
On the other hand, the HH neuron belonging to
the type II has the discontinuous
$f_{\rm o}-I_{\rm i}$ relation at the critical
current, $I_{\rm c}$, above which it shows the
oscillation with a narrow range of $f_{\rm o}$ (see Fig.1(a)).
For {\it inhibitory} (hyperpolarizing) input currents,
the HH neuron may bring about a firing by the
so-called rebound process against our intuition 
\cite{Perkel74}, while the IF neuron cannot.
This inhibitory rebound is realized in
some non-linear dynamical neuron models 
like Fitzgh-Nagumo model.
 
Quite recently, the present author \cite{Hasegawa99} 
has investigated the
responses of the HH model to various types of
{\it spike-train} inputs with constant, chaotic and
stochastic  interspike intervals (ISI):
Ref.\cite{Hasegawa99} is referred to as I in this paper.
Our calculation in I shows that behavior of the variability
to stochastic ISI inputs of the HH model is rather different
from that obtained based on the 
IF model \cite{Feng98}\cite{Marsalek97}.
This is consistent with the recent calculation of
Brown, Feng and Feerick, \cite{Brown99}, who show
a stronger dependence of the variability on the level
of inhibitory inputs in the IF neurons than 
in the HH neurons.

It is the purpose of the present paper to elucidate
the origin of the difference of the responses of
IF and HH models to spike-train inputs, investigating
whether the IF model may be an adequate substitute of 
the HH model.
In order to clarify the point, we first employ the
spike-train inputs with simple constant ISIs, 
and then the stochastic ISI inputs
to get some insight to the controversial variability problem
initiated by Softky and Koch \cite{Softky92}.

Our paper is organized as follows:
In the next sec.II, we describe a simple neuron 
model adopted for our numerical calculation.
In sec.III, we investigate the response of our system to
inputs with the constant ISI.
Stochastic inputs with the Gamma distribution are
treated in sec.IV.
The final section VI is devoted to conclusion and discussion.
Although some of the calculated results for the HH neuron
have been published in I, we include them for the
completeness of the present paper.
\vspace{1.0cm}
\noindent
{\bf 2. Adopted model}

We adopt a simple system consisting of a neuron and a synapse; 
the former is described by the HH or IF model and
the latter by the alpha function.
We will investigate the response of our neuron when
spike-train inputs are applied through the synapse.

\vspace{0.5cm}
\noindent
{\it 2.1 Hodgkin-Huxley model}

The HH model is described by the non-linear coupled differential
equations for the four variables, $V$ for the membrane potential,
and $m, h$ and $n$ for the gating variables of Na and
K channels, and it is given by \cite{Hodgkin52}
\begin{equation}
\bar{C} d V/d t = -g_{\rm Na} m^3 h (V - V_{\rm Na})
- g_{\rm K} n^4 (V - V_{\rm K}) 
- g_{\rm L} (V - V_{\rm L}) + I_{\rm i},
\end{equation}
\begin{equation}
d m/d t = - (a_m + b_m) \: m + a_m,
\end{equation}
\begin{equation}
d h/d t = - (a_h + b_h) \: h + a_h,
\end{equation}
\begin{equation}
d n/d t = - (a_n + b_n) \: n + a_n.
\end{equation}
Here  the reversal potentials of Na, K channels and leakage are  
$V_{\rm Na} = 50$ mV, $V_{\rm K} = -77$ mV and 
$V_{\rm L} = -54.5 $ mV;
the maximum values of corresponding conductivities are
$g_{\rm Na} = 120 \; {\rm mS/cm}^2$, 
$g_{\rm K} = 36 \; {\rm mS/cm}^2$ and
$g_{\rm L} = 0.3 \; {\rm mS/cm}^2$; the capacity of the membrane is
$\bar{C} = 1 \; \mu {\rm F/cm}^2$; detailed expressions
for $a_m$, $b_m$ {\it et al.} are presented
in Refs.\cite{Hodgkin52} \cite{Hasegawa99} \cite{Park96} .

\vspace{0.5cm}
\noindent
{\it 2.2 Integrate-and-Fire model}

We adopt the IF model which
may include the absolute refractory period. 
The dynamics of the membrane potential, $V$, 
and the phenomenologically introduced variable, $p$,
is described by
\begin{equation}
C \: dV/dt = -g\:(1 + p\:a)(V - V_{\rm r} - p V_{\rm d}) + (1 - p) I,
\end{equation}
\begin{equation}
dp/dt = (-1/\tau_{\rm p})\:[p - \Theta(p - w)],
\end{equation}
with $w = (V_{\rm t} - V)/(V_{\rm t} - V_{\rm r})$, 
$a = \tau_{\rm r}/\tau_{\rm m} -1$ and
$\Theta (x)$ is the Heaviside step function.
Here $C$ is the capacitance and $g \:(=C/\tau_{\rm m})$ 
the conductance of the membrane; $\tau_{\rm m}$ 
is the life time of membrane potential and
$\tau_{\rm r}$ refractory period; $V_{\rm r}$  and $V_{\rm t}$ 
are the reset and threshold potentials, respectively.
In the limit of $\tau_{\rm r} \rightarrow 0$ and 
$\tau_{\rm p} \rightarrow 0 $, our IF model 
given by Eqs.(5) and (6) is equivalent to the 
conventional IF model \cite{Maass97},
in which the reset condition is given by
$V(t_{{\rm o}m}^{-})=V_{\rm t}$ and 
$V(t_{{\rm o}m}^{+})=V_{\rm r}$
at $t_{{\rm o}m}$ , the firing time of $V$ $(m=0,1,..)$.

The variable $p$ is zero in the active period and it is one in 
the refractory period.
The rapid transient between the two states of $p$ is 
given by Eq.(6) with a small $\tau_{\rm p}$. 
In the active period ($p = 0$), Eq.(5) becomes
\begin{equation}
dV/dt = - (1/\tau_{\rm m})(V - V_{\rm r}) +  I/C,
\end{equation}
which is just the same as that given in the conventional
leaky IF model \cite{Maass97}.
On the other hand, in the refractory period ($p = 1$) Eq.(5) becomes
\begin{equation}
dV/dt = - (1/\tau_{\rm r})(V - V_{\rm r} - V_{\rm d}),
\end{equation}
where $\tau_{\rm r} = \tau_{\rm m}/(1 + a)$. 
Equation (8) shows that the reset of $V$ is accomplished
by the deriving potential, 
$V_{\rm r}+V_{\rm d}$,
with the time constant $\tau_{\rm r}$ which is smaller 
than $\tau_{\rm m}$.

The advantages of the present model given by Eqs.(5) and (6)
are that it includes the refractory period and that it
automatically resets the membrane potential.
The model similar to ours was previously proposed 
by Horn and Opher\cite{Horn97}.
The physical meaning of their model is, however, not
transparent and their $\alpha$ term, which corresponds to 
our $V_d$ term in Eq.(6), improperly 
persists in the active period.
It is crucial to reject this term in the active period (Eq.(7))
for a discussion of the input-output response of the IF neuron.

The bold curve in Fig.1 expresses the $f_{\rm o}-I_{\rm i}$
relation of the HH neuron, showing the discontinuous transition
at $I_{\rm c} = 6.3 \: \mu {\rm A/cm}^2$.
It is rather difficult to choose the parameters of
the IF model so as to reproduce the $f_{\rm o}-I_{\rm i}$ 
relation of the HH neuron because the IF and
HH neurons belong to the different type neurons,
although the exact choice of parameter values is not crucial.
After several tries \cite{Comment1},
we have determined to employ the 
following parameters for our numerical calculations:
$V_{\rm r} = -75$ mV, $V_t = -55$ mV, $V_d = -10$ mV, 
$\tau_m = 20$ msec \cite{Brown99}, 
$\tau_p = 0.02$ msec 
and $C = 4 \mu {\rm F/cm}^2$.
As for the refractory period, $\tau_{\rm r}$, we adopt the 
two choices; neurons with 
$\tau_{\rm r} = 0.1$ and 2.0 msec are referred to 
as the {\it IF0} and {\it IF1 neurons}, respectively. 

After a simple calculation, we get the $f_{\rm o}$ dependence 
on $I_{\rm i}$ of IF neurons, given by
\begin{equation}
1/f_{\rm o} = T_{\rm o} = \tau_{\rm m} \: {\rm ln} 
( \frac{I_{\rm i}}{I_{\rm i}-I_{\rm c}} )
+ \tau_{\rm r} \: {\rm ln} 
[ \frac{V_{\rm d}-(V_{\rm t}-V_{\rm r})}{V_{\rm d}} ] 
+ O(\tau_{\rm p}),
\end{equation}
where the critical current, $I_{\rm c}$, is given by
\begin{equation}
I_{\rm c} = (C/\tau_{\rm m})\: (V_{\rm t} - V_{\rm r}),
\end{equation}
leading to $I_{\rm c} = 4 \; \mu {\rm A/cm}^2$.
Dashed and solid curves in Fig.1 express
$f_{\rm o}$ for IF0 and IF1 neurons, respectively.

Figure 1(b) shows the examples of the self-excited 
oscillations of the membrane potentials
of the HH, IF0 and IF1 neurons. For IF neurons
we plot $V + c \: p(1-p)$  instead of $V$:
the $c$ term ($c =350$) yields the spiky contribution. 

\vspace{0.5cm}
\noindent
{\it 2.3 Synaptic inputs}

We consider the delta-function-type spike-train input expressed by
\begin{equation}
U_{\rm i}(t) = V_{\rm a} \: \sum_n  \: \delta (t - t_{{\rm i}n}),
\end{equation}
where $V_{\rm a}$ is the magnitude of the action potential,
and the firing time $t_{{\rm i} n}$ for arbitrary $n$
is assumed to be recurrently defined by
\begin{equation}
t_{{\rm i}n+1} = t_{{\rm i}n} 
+ T_{{\rm i}n}(t_{{\rm i}n}),
\end{equation}
\begin{equation}
t_{{\rm i}1} = 0,
\end{equation}
ISI of input spike, 
$T_{{\rm i}n}$, being
generally a function of a given time $t_{{\rm i}n}$.  

It has been reported that biological synapses exhibit temporal dynamics
during neuronal computations \cite{Abbott97} \cite{Tsodyks98}.
We, however, treat the synapse as a static unit for a simplicity of
our calculation. 
The spike train given by Eq.(11) is injected
through the synapse, yielding the 
postsynaptic current $I_{\rm i}$ given by
\begin{equation}
I_{\rm i}(t) 
= A_{\rm syn} \: 
\sum_n \: \alpha (t - t_{{\rm i}n}), 
\end{equation}
where $A_{\rm syn}=g_{\rm syn} \: (V_{a}-V_{\rm syn})$,
$g_{\rm syn}$ and $V_{\rm syn}$ are 
the synaptic conductance and reversal potential,
respectively, and
the alpha function, $\alpha(t)$, 
is defined by \cite{Hasegawa99}\cite{Park96}
\begin{equation}
\alpha(t) = (t/\tau_{s}) \; e^{-t/\tau_s} \:  \Theta(t),
\end{equation}
$\tau_s$ being the time constant relevant to the 
synapse conduction.
The positive and negative $A_{\rm syn}$ stand for
the excitatory and inhibitory couplings, respectively.
When $T_{{\rm i}n} \gg \tau_{s}$,
Eqs.(14) and (15) yield  pulse currents with the 
maximum value of
$I_{\rm i}^{\rm max}=0.368 \: A_{\rm syn}$
at $t=T_{{\rm i}n}+\tau_{\rm s}$ and with the half-width of 
$2.45 \: \tau_{\rm s}$.
On the contrary, 
when $T_{{\rm i}n} \siml \tau_{s}$ the temporal summation
of input currents is realized because an input pulse comes
before the current induced by its preceding pulse 
is not attenuated. We assume 
$\tau_s = 2$ msec
and treat $A_{\rm syn}$ as an adjustable parameter.

When the membrane potential $V$ oscillates,
it yields the spike-train output, which may be expressed by
\begin{equation}
U_{\rm o}(t) = V_a \: \sum_m \: \delta (t - t_{{\rm o}m}),
\end{equation}
in a way similar to Eq.(11), where $t_{{\rm o}m}$ is defined 
as the time when the membrane potential $V(t)$ crosses $V_{z}=0$ mV
from below. The output ISI is defined by
\begin{equation}
T_{{\rm o}m} = t_{{\rm o}m+1} - t_{{\rm o}m}.
\end{equation}

Differential equations given by Eqs.(1)-(4) for the HH neuron
(or Eqs.(5) and (6) for the IF neuron) including
the external current given by Eqs.(11)-(15) are solved 
by the forth-order Runge-Kutta method  
with the integration time step of 0.01 msec.
The calculations are performed for 2 sec in the constant-ISI case
and for 20 sec in the stochastic ISI case.
If ISI of spike-train input or output
is about 10 msec in the latter case, the size of its sample 
is about 2000.
Although this figure is not sufficiently large for statistics
of ISI data, we hope an essential ingredient will be clarified
in our numerical investigation.
Hereafter time ($T_{{\rm i}n}$ etc.), voltage ($V$ etc.) and 
current ($I_{\rm i}$, $A_{\rm syn}$ etc.) are
expressed in units of msec, mV and $\mu{\rm A/cm}^2$,
respectively unless otherwise specified.
 
%
\vspace{1.0cm}
\noindent
{\bf 3. Constant-ISI inputs} 

\noindent
{\it 3.1 HH neurons}
%

Let us first consider the HH neuron which receives 
spike-train inputs given by Eqs.(11)-(15)  
with the positive $A_{\rm syn}$ and constant $T_{{\rm i}n}$.
The solid curve in Fig.2(c) expresses the time course of the
membrane potential, $V$, for the excitatory postsynaptic
current, $I_{\rm i}$, with 
$A_{\rm syn} = 40$ shown
by the solid curve in Fig.2(b), which is induced by 
an applied spike-train input, $U_{\rm i}$, 
with $T_{{\rm i}n}=20$ msec as depicted in Fig.2(a)
(for time courses of potentials for inputs 
with $T_{{\rm i}n}=10$ msec, see Fig.5 of I).
For this spike-train input of $T_{{\rm i}n}=20$ msec,
we get a regular spike-train output with $T_{{\rm o}m}=20$ msec.
This is in contrast with the case of spike-train input of
$T_{{\rm i}n}=10$ msec, in which 
output spike-train is phase locked with the ratio of $4 : 3$, 
oscillating with a long cycle of
40.00 msec (=11.25 + 12.36 + 16.39) = $4 \: T_{{\rm i}n}$, 
where 11.25, 12.36 and 16.39 are the values of output ISIs.

Solid and dashed curves in Fig.3(a) show the average 
($\mu_{\rm o}$)
and root-mean-square (RMS, $\sigma_{\rm o}$) values of 
$\{ T_{{\rm o}m} \}$
for input ISI of $T_{{\rm i}n}=10$ msec 
when $A_{syn}$ is changed .
Filled circles denote the distribution of $\{ T_{{\rm o} m} \}$
for a given $A_{\rm syn}$.
For example, in the case of $A_{\rm syn} = 40$,
we get $T_{{\rm o} m} $ = 11.25, 12.36 and 16.39 msec as
mentioned above.
For $A_{\rm syn} > 56$, 
ISI of the output is the same as
that of input, then the coefficient defined by 
$k = \mu_{\rm o}/\mu_{\rm i} $ is unity.
When $A_{\rm syn}$ is decreased, we get $k$ larger than unity, and
$k = 2$ for $8 < A_{\rm syn} < 28$.  
Below $A_{\rm syn} = 8$, 
we have no output spikes.

Solid and dashed curves in Fig.3(b)  express $\mu_{\rm o}$
and $\sigma_{\rm o}$ when the average of input ISI
$(\mu_{\rm i})$ is changed with 
$A_{\rm syn} = 40$.  
We obtain $k = 1$ for $\mu_{\rm i} \simg 12$ msec, 
$k=2$ for $6 \siml \mu_{\rm i} \siml 8$ msec,
and $k = 3$ for $\mu_{\rm i} = 4$ msec: 
otherwise  $k$ is non-integer. 
We obtain no output ISIs of $T_{{\rm o}m} < 10$ msec
for $T_{{\rm i}n} < 12 $ msec, which is due to
the low-pass filter character of the HH neuron \cite{Hasegawa99}.

Next we discuss the inhibitory input case.
The dash curve in Fig.2(c) expresses the time course
of membrane potential with $T_{{\rm o}m}=20$ msec
when we apply the inhibitory postsynaptic
current with $A_{\rm syn}=-40$
and $T_{{\rm i}n}=20$ msec, which 
works to hyperpolarize
the membrane potential, $V$. 
When the postsynaptic current given by the alpha function
(Eq.15) decreases and vanishes, $V$ changes to restore
to the rest level and it crosses
the threshold to yield an action potential 
with a delay of about 15 msec.
Thus the HH neuron can fire even for the inhibitory input
by the rebound process \cite{Perkel74}.
This process, however, requires an appreciable periods:
when the input ISI is less than about 15 msec, the HH neuron cannot
fire because the next inhibitory spike is applied before
the hyperpolarized membrane potential crosses
the threshold level. 

By changing $A_{\rm syn}$ and $\mu_{\rm i}$, 
we perform similar calculations,
whose result is summarized in Fig.4(a). It expresses the phase diagram 
showing the region where the {\it integer} $k$ is obtained in the
$\mu_{\rm i}-A_{\rm syn}$ space.
Note that between the integer $k$'s, we have non-integer solutions;
for example, in the case of $\mu_{\rm i}=10$ msec
and $A_{\rm syn}=40$, 
we get $k=1.33$ between $k = 1$ and 2.
We notice for positive $A_{\rm syn}$ 
that there is the wide $k=1$ region and that 
the regions of larger $k$ appear at the left side of
the $k=1$ region.
For negative $A_{\rm syn}$, the HH neuron fires for input
with $\mu > 15$ msec, for which we get the states 
of $k=1$ and $k=2$.


%

\vspace{0.5cm}
\noindent
{\it 3.2 IF neurons}

Now we consider the IF neuron to which the constant-ISI 
input is applied.  
The solid (dashed) curve in Fig.2(d) shows the time course
of the membrane potentials of the IF1 neuron for 
excitatory (inhibitory) inputs.
When the excitatory input with $T_{{\rm i}n}=20$ msec is applied,
its membrane potential is depolarized to cross the threshold, and
it emits the spike-train output with $T_{{\rm o}m}=20$ msec.
When the inhibitory input is applied, its membrane potential
is hyperpolarized. After postsynaptic current vanishes,
$V$ changes to restore to the rest level but cannot cross
the threshold level. The behavior of $V$ of the IF1 neuron
is in strong contrast with that of the HH neuron shown 
by the dashed curve in Fig.2(c).
Thus the IF1 (and IF0) neuron is not excitable for
inhibitory spike-train inputs.

Figure 5 shows the time
courses of $U_{\rm i}(t)$ and $V(t)$ of the IF1 neuron 
for $T_{{\rm i}n} = 10$ msec
with $A_{\rm syn}$ = 40, 64 and 120.
For $A_{\rm syn}=64$, IF1 neurons regularly emit spike trains
of $T_{{\rm o}m}=10$ msec with a delay of about 3 msec.
For $A_{\rm syn}=40$, the output ISI becomes larger than 10 msec,
because it can emit output pulse after integrating small input
signals by virtue of the type-I neuron although a single input
pulse is insufficient to trigger the output pulse.
On the contrary, for a large $A_{\rm syn}=120$, a single input 
pulse make the IF1 to emit irregularly multiple pulses with
$T_{{\rm o}m}$ smaller than $T_{{\rm i}n}$.
The time course of the membrane potential of the IF0 neuron is
ostensibly similar to that of the IF1 neuron (not shown).

The $A_{\rm syn}$ dependence of the responses of 
the IF0 and IF1 
neurons to inputs with $\mu_{\rm i} = 10 $ msec is shown 
in Fig.6(a) and 7(a), respectively.
Solid  (dashed) curves express $\mu_o$ ($\sigma_{\rm o}$) 
and filled circles denote $\{ T_{{\rm o}m} \}$ for 
a given $A_{\rm syn}$.
For the IF0 neuron, we get the integer values of
$k=1$ for  $51 \siml A_{\rm syn} \siml 58$, 
$k=2$ for $A_{\rm syn} = 32$,
$k=3$ for $A_{\rm syn}=26$ and
$k=4$ for $A_{\rm syn}=22$.
On the contrary, for the IF1 neuron
we get the integer values of 
$k=1$ for  $54 \siml A_{\rm syn} \siml 83$, 
$k=2$ for $32 \siml A_{\rm syn} \siml 37$,
$k=3$ for $A_{\rm syn}=26$ and
$k=4$ for $A_{\rm syn}=22$.
When we compare these results with that of HH neurons 
(Figs. 3(a) and (b)),
we notice that $k \simg 3$ is realized for small $A_{\rm syn}$
and that $k < 1$ exists at large $A_{\rm syn}$.

Figures 6(b) and 7(b) show the $\mu_{\rm i}$ 
dependence of $\mu_{\rm o}$ and
$\sigma_{\rm o}$ of the IF0 and IF1 neurons, respectively,
with $A_{\rm syn} = 64$.
We get $\mu_{\rm o} 
= \mu_{\rm i} \;(k=1)$ for
$\mu_{\rm i} > 15$ ($\mu_{\rm i} > 6$) msec in the IF0 (IF1) neuron.
When $A_{\rm syn}$ is reduced, we obtain the states with 
larger $k$.  For example, for $A_{\rm syn}=40$,  
we get $k=2$ with 
$16 < \mu_{\rm i} < 22$ ($12 < \mu_{\rm i} < 22$) msec 
in the IF0 (IF1) neuron; the $k=1$ state is 
not available (not shown).

From calculations by changing $A_{\rm syn}$ and $\mu_{\rm i}$,
we obtain the phase diagrams  of the integer $k$ values
in the $\mu_{\rm i}-A_{\rm syn}$ space for the IF0 and IF1 neuron,
which are shown in Figs.4(b) and (c), respectively.
Because IF0 and IF1 neurons cannot fire for inhibitory inputs,
results only for the positive $A_{\rm syn}$ are shown.
A comparison of them with the corresponding
phase diagram of the HH neuron (Fig.4(a)) 
shows 

\noindent
(1) the $k=1$ region, particularly of the IF0 neuron,
is greatly reduced,   

\noindent
(2) although the $k \geq 2$ region in the
HH neuron appears at the left side of the $k=1$ phase, 
such phases in IF0 and IF1 neurons appear below the $k=1$ phase,

\noindent
(3) the phase with $k < 1$ appears for IF0 and IF1 neurons 
although it does not exist for the HH neuron
in the parameter range shown in the Fig.4(a), and

\noindent
(4) an inclusion of the refractory period in the IF1 neuron
widens the $k=1$ region.

\vspace{1.0cm}
\noindent
{\large\bf 4. Stochastic-ISI inputs} 

The ISIs of spike-train input, $T_{{\rm i} n}$, 
in Eq.(12) are assumed to be
independent random variables with the Gamma probability
density function given by
\begin{equation}
P(T) = s^r \;\; T^{r - 1} \;\; e^{- s T}/\; \Gamma(r)
\end{equation}
for which we get $\mu_{\rm i} =  r/s$, and
$\sigma_{\rm i} = \sqrt{ r}/s$,
$\Gamma \;(r)$ being the gamma function.
For a later purpose, we define the dimensionless 
variability  given by
\begin{equation}
c_{\rm v \lambda} = \sigma_{\lambda}/\mu_{\lambda}  
\:\:\: \:\:\: \:\:\:
\mbox{($\lambda$ = i and o)}, 
\end{equation}
for input ($\lambda={\rm i}$) and 
output ISIs ($\lambda={\rm o}$),
from which we get $c_{\rm vi} = 1/\sqrt{r}$ 
for the Gamma-distribution inputs.
For $r = 1$ in Eq.(18) we recover 
the exponential distribution 
($c_{\rm vi}$ = 1) and a Poisson distribution for 
the number of spikes in a given time interval.
In the limits of $r \rightarrow \infty$ 
and $s \rightarrow \infty$ with keeping 
$\mu_{\rm i} = r/s $ fixed, Eq.(18) reduces to
$P(T) = \delta(T - \mu_{\rm i}) $, the constant ISI
with $\mu_{\rm i} = T$ and $c_{\rm vi} = 0$.

The spike-train input created by the Gamma-distribution 
generator is applied to our neural system.
Calculations are performed by changing 
$A_{\rm syn}$ or $\mu_{\rm i}$ by keeping the value of 
$c_{\rm vi}$ fixed.
Note that because the number of our sample of
input ISI is not sufficiently large, the obtained
$c_{\rm vi}$ fluctuates around the intended values.

\vspace{0.5cm}
\noindent
{\it 4.1 HH neurons}

The histogram in Fig.8(a) shows the distribution of 
excitatory input ISIs ($A_{\rm syn}=40$)
with $c_{\rm vi}$ = 0.40 and $\mu_{\rm i} = 10$ msec, 
which leads to output ISIs with the distribution
shown in Fig.8(b)
($c_{\rm vo}$ = 0.25 and $\mu_{\rm o}$ = 14.84 msec).
Output histogram in Fig.8(b) has no distributions at
$T_{{\rm o}m} < 10$ msec, which arises from 
the low-pass filter character
of the HH neuron as shown in Fig.3(b).

Figures 9(a) and (b) show the results of 
$\mu_{\rm o}$, $\sigma_{\rm o}$
and $c_{\rm vo}$ as a function of $A_{\rm syn}$ for inputs
with $c_{\rm vi} = 0.4$ and 1.0, respectively 
($\mu_{\rm i}=10$ msec). Comparing them 
with the results for the constant ISI ($c_{\rm vi} = 0$)
shown in Fig.3(a), 
we note that for larger $c_{\rm vi}$, values of 
$\mu_{\rm o} > 20$ msec
are much accumulated in the region of smaller $A_{\rm syn}$ region.
This is due to the integration character of neurons 
for weak inputs and it is also realized in the IF neuron 
(Figs.10(a) and 11(a)).  

The dependence of output ISIs on $\mu_{\rm i}$ for 
$c_{\rm vi} = 0.4$ and 1.0
are shown in Fig.9(c) and (d), respectively.
We note that the structure
at $\mu_{\rm i} < 10$ seen in the case of $c_{\rm vo}=0$ (Fig.3(b),
disappears because of the randomness in inputs.
Although $\mu_{\rm o} \sim \mu_{\rm i}$ for large $\mu_{\rm i}$,
we get $\mu_{\rm o} > 10$ msec for small $\mu_{\rm i}$,
since the HH neuron plays as the low-pass filter.
For $c_{\rm vi} = 1.0$, we get 
$\mu_{\rm o} \sim (\mu_{\rm i}+10)$ whereas
$\sigma_{\rm o} \sim \mu_{\rm i}$, 
which yield an increase in $c_{\rm vo}$
as increasing $\mu_{\rm i}$.

Figures 9(e) and (f) express the $\mu_{\rm i}$ dependence
of output ISIs for inhibitory inputs with $c_{\rm vi}=0.4$ and 1.0,
respectively. Their comparisons with the result for $c_{\rm vi}=0.0$
(Fig.3 (c)) show that the HH neuron may fire 
for stochastic inputs with $\mu_{\rm i} < 15$ msec.
It is interesting to note from a comparison 
of Fig.9(c) (Fig.9(d)) with Fig.9(e) (Fig.9(f))
that the $\mu_{\rm i}$ dependence
for inhibitory inputs is quantitatively
similar to that for excitatory inputs. 

\vspace{0.5cm}
\noindent
{\it 4.2 IF neurons}

Histograms in Figs.8(c) and (d) show the distributions of
output ISIs of the IF0 neuron and those of 
the IF1 neuron, respectively, for excitatory inputs of
$c_{\rm vi} = 0.4$ and $\mu_{\rm i} = 10$ msec
whose distribution is plotted in Fig.8(a). 
Solid and dashed histograms in Fig.8(c) and (d) express the
results of $A_{\rm syn} = 40$  and 64, respectively.
The input histogram has a peak at about 
$T_{\rm in} = 10$ msec, as expected.
On the contrary, output ISIs of IF0 and IF1 neurons
for $A_{\rm syn} = 64$ have peaks at lower values of ISI.
This is understood from Figs.6(a) and 7(a) which show that
$k < 1$  for $\mu_{\rm i} < 16 \; (12)$
msec in the IF0 (IF1) neuron.
Furthermore, for the IF1 neuron, the refractory period
makes the neuron to emit no outputs for  
$T_{{\rm o}m} < 4 $ msec.

Figures 10(a) and (b) (11(a) and (b)) show output results
for excitatory inputs with $c_{\rm vi}=0.4$ and 1.0, respectively, 
of the IF0 (IF1) neuron as a function of $A_{\rm syn}$.
As $c_{\rm vi}$ is increased, large $\mu_{\rm o}$ values 
are accumulated in the region with a small $A_{\rm syn}$.
Although this phenomenon is realized also
in the HH neuron (Fig.9(a) and (b)), the $\mu_{\rm o}$ value
is 100 msec at most in the HH neuron
while the maximum $\mu_{\rm o}$ 
value exceeds 1000 msec in the IF neurons.
The difference in $\mu_{\rm o}$ and $\sigma_{\rm o}$ between
the IF0 and IF1 neurons become smaller for larger
$c_{\rm vi}$, although $c_{\rm vo}$ of the IF0 neuron
is always larger than that of the IF1 neuron.

The $\mu_{\rm i}$ dependence of the output ISIs
for excitatory inputs with $c_{\rm vi}=0.4$ and 1.0
of the IF0 (IF1) neuron are shown in
Fig.10(b) and (d) (11(b) and (d)), respectively.
With increasing $\mu_{\rm i}$, both $\mu_{\rm i}$ and 
$\sigma_{\rm i}$ increase, but
$c_{\rm vo}$ tends to saturate.  
For larger $c_{\rm i}$, we get larger $c_{\rm o}$, as expected.
It is interesting to compare the results of 
the IF neurons with those of the HH neuron.
For inputs with $c_{\rm i}=0.4$ and $\mu_{\rm i}=10-30$ msec, 
we obtain $c_{\rm vo}= 0.25-0.40$, 0.47-0.77 and 0.40-0.45
for the HH, IF0 and IF1 neurons, respectively.
Similarly, for inputs with $c_{\rm i}=1.0$ and 
$\mu_{\rm i}=10-30$ msec, 
we get $c_{\rm vo}= 0.56-0.86$, 1.01-1.13 and 0.85-0.96
for the HH, IF0 and IF1 neurons, respectively.
We note that the relation:

\begin{equation}
c_{\rm vo}^{\rm HH} < c_{\rm vo}^{\rm IH1} 
< c_{\rm vo}^{\rm IF0} \sim c_{\rm vi},
\end{equation}
holds in our calculation, related discussion being
given in the next section.

\vspace{1.0cm}
\noindent
%
{\bf 5. Conclusion and discussion}

Since Softky and Koch \cite{Softky92} 
reported a large $c_{\rm vo} \; (\sim 0.5-1.0)$ in cortical neurons 
in visual V1 and MT of monkeys,
it has been controversial 
how to understand the large variability of neurons to stochastic 
inputs \cite{Softky92}\cite{Shadlen94}-\cite{Gutkin98}. 
There have been much discussions on this subject using the 
IF model.
Some theoretical studies show that IF models lead to small
$c_{\rm vo}$ because an integration of a large number of
random inputs works to reduce the 
variability \cite{Abbott97}\cite{Softky92}.
On the other hand, other studies have shown that
IF neurons may yield an appreciable value of 
$c_{\rm vo}$ \cite{Feng98}\cite{Brown99}.

Figure 12 shows the $c_{\rm vi}-c_{\rm vo}$ plot of ISI 
data having reported for HH (Figs.3(b), 9(c) and 9(d)),
IF0 (Figs.6(b), 10(c) and 10(d)) 
and IF1 neurons (Figs.7(b), 11(c) and 11(d))
and of new results calculated
for $c_{\rm vi}=0.7$.  Open squares, filled triangles 
and circles denote the results of the HH, IF0 and IF1 neurons, 
respectively.
We note the variability of output ISIs may be large 
and nearly the same as that of input ISIs 
in HH and IF neurons.
Scattered values of $c_{\rm vo}$ for a given $c_{\rm vi}$ 
mean that $c_{\rm vo}$ depends not
only on $c_{\rm vi}$ but also strongly on $\mu_{\rm i}$
(and $A_{\rm syn}$).
For example, when the $\mu_{\rm i}$ is
varied from 2 to 30 msec in the IF0 neuron, we get 
$c_{\rm vo}= 0.20-0.48$ for $c_{\rm vi}=0.4$,
$c_{\rm vo}= 0.40-0.84$ for $c_{\rm vi}=0.7$ and
$c_{\rm vo}= 0.64-1.18$ for $c_{\rm vi}=1.0$.
Our calculation reconciles the dispute among
the earlier calculations
yielding small  \cite{Abbott97}\cite{Softky92}
and large $c_{\rm vo}$\cite{Feng98}\cite{Brown99}\cite{Troyer97}
based on the IF model;
the difference in the calculated $c_{\rm vo}$ may be due
to the difference in the parameters adopted in their
calculations.  

It should be noted in Figs.9, 10 and 11 that
the variability of output ISIs generally increases 
with increasing input ISIs, which is in agreement
with the biological data (Fig.3 of Ref.\cite{Softky93}).
When the input ISI is small compared to the characteristic
integration time, the neuron acts as an integrator yielding 
a small $c_{\rm vo}$, while when the reserve is true,
the neuron play a role of the coincidence detector with
a fairly large $c_{\rm vo}$.

To summarize, we have performed numerical calculations
of the responses of the HH and IF neurons
to inputs with the constant and stochastic ISIs,
to make a comparison
between them,
after we had chosen the parameters such as 
for the IF model to mimic the time dependence of
the membrane potentials of the HH neuron (Fig.1(a)). 
Our calculations have shown the followings:

\noindent
(i) the HH neuron can fire for both excitatory and
inhibitory inputs whereas the IF neuron only 
for excitatory inputs,
 
\noindent
(ii) responses of the HH and IF neurons 
show the complicated behavior to spike-train inputs
even with the simple constant ISIs, for which
$\mu_{\rm o}$ (or $k=\mu_{\rm o}/\mu_{\rm i}$) is 
generally functions of
$\mu_{\rm i}$, $c_{\rm vi}$ and $A_{\rm syn}$,

\noindent
(iii) the response to constant ISIs of the IF neurons is 
rather different
from that of the HH neuron: the $k=1$ region of the IF neuron,
in which neurons properly respond to inputs, 
is much narrower than that of the HH neuron (Fig.4),

\noindent
(iv) the variability of IF and HH neurons to
stochastic ISIs shown in Fig.12,
may be large and follow the relation given by Eq.(20),

\noindent
(v) the $A_{\rm syn}$ dependence of the variability
of the IF neuron is stronger than that of the HH neuron,
and

\noindent
(vi) an inclusion of the refractory period
in IF1 depresses the $c_{\rm vo}$ values and improves
to some extent its response,  widening the $k=1$ region.

The difference in the item (i) arises from the fact that
the HH model has the rebound process against the 
hyperpolarized membrane potential
while the IF model does not.
In order to make the item (ii) more concrete, we show
in Figs.13(a) and (b), the $f_{\rm i}-f_{\rm o}$ plots
of the HH and IF0 neurons, respectively,
where $f_{\rm \lambda}\; (=1/\mu_{\rm \lambda })$ is
the mean frequency of input $(\lambda={\rm i})$
and output ISIs $(\lambda={\rm o})$.
Saturating functions with the maximum frequency of
$f_{\rm o}^{\rm max} \sim 100$ Hz given by
the $f_{\rm i}-f_{\rm o}$ plots of the HH neuron 
(Fig.13(a)), are similar to the  
sigmoidal function of $g(x)=(1+{\rm tanh} \: x)/2$
adopted in the formal rate-coding  models \cite{Hopfield82}.
However, monotone increasing functions given by 
those of the IF0 neuron (Fig.13(b)) are 
quite different from $g(x)$.
The $f_{\rm i}-f_{\rm o}$ plots of the IF1 neuron
express also increasing functions but saturate with 
the large maximum frequency of 
$f_{\rm o}^{\rm max} \sim 500$ Hz (not shown). 
The main origins yielding the differences cited
in the items (iii)-(vi) are
(1) the difference of the $f_{\rm o}-I_{\rm i}$ relation of 
the type-I IF neuron from that of the type-II HH neuron and
(2) the difference in the refractory period.
The continuous $f_{\rm o}-I_{\rm i}$ relation in the IF neuron
yields a large $k \;(\geq 3$)
after integrating small inputs for a small $A_{\rm syn}$.
The IF models with
the vanishing or small refractory
periods emits the output ISI with
$k < 1$.  Then the $k=1$ region in the
phase diagram of the IF neuron is much reduced compared with
that of the HH neuron (Fig.4).
The item (iv) is consistent with earlier calculations
using the IF \cite{Feng98}\cite{Brown99}\cite{Troyer97} 
and HH models \cite{Brown99}.
Brown, Feng and Feedick \cite{Brown99} show
that the variability of the IF neuron
has a stronger dependence on the number of synaptic 
inputs, $N_{\rm s}$, than that of the HH neuron.
Since their $N_{\rm s}$ is expected to correspond
to our $A_{\rm syn}$ in a crude sense,
the item (v) is consistent with their result.  
They \cite{Brown99} also claim that an inclusion 
of the absolute refractory period in the IF neuron 
decreases the variability and that it
{\it increases} the disparity between the results for
the IF and HH neurons.  The former agrees with 
our item (vi) but the latter does not.
Gutkin and Ermentrout \cite{Gutkin98} 
predict based on the Morris-Lecar (ML) model
that the variability of type-I ML neuron
is larger than that of type-II ML one,
which is supported by our calculations (Eq.20)).
However, their claim that the type-II neuron
yield only a small $c_{\rm vo}$, cannot be applied to 
the HH model.

The HH neuron responds to static
and spike inputs differently from the IF neuron.
The HH neuron shows the complex behavior
not shared with an IF neuron. 
The chaotic oscillation induced 
by an applied sinusoidal currents
\cite{Aihara84}\cite{Matsumoto84} and the 
firing by the inhibitory rebound \cite{Perkel74}
are never realized in the single IF neuron.
Our calculations have shown that even for
the excitatory inputs the IF model
may be a good substitute of the HH model only 
within the limited parameters.
There are many experimental and theoretical evidences
showing that reciprocally inhibitory neurons
play important roles in real systems such as
Hippocampus and thalamus \cite{Hippo}, for which the IF model
cannot be used.
We should mind advantage and disadvantage 
of the IF neuron in modeling
biological neural systems.

\vspace{1.0cm}
\noindent{\bf Acknowledgment}  \par
\vspace{0.2cm}

This work is partly supported by
a Grant-in-Aid for Scientific Research from the Japanese 
Ministry of Education, Science and Culture.

\newpage
%

%

\newpage
\noindent{\large\bf  Figure Captions}   \par
\vspace{0.5cm}

\noindent
{\bf Fig.1} 
(a) The $f_{\rm o}-I_{\rm i}$ plot of the HH (bold solid curve), 
IF0 (dashed curve) and IF1 neurons (solid curve), and
(b) the self-excited oscillations of the HH, IF0 and IF1 neurons,
scales for IF0 and IF1 neurons being shifted by 200 and 400 mV,
respectively.
\vspace{0.5cm}

\noindent
{\bf Fig.2} 
The time courses of
(a) the input spike train, $U_{\rm i}$, 
with $T_{{\rm i}n}=20 $ msec,
(b) the postsynaptic currents, $I_{\rm i}$,
and (c) the membrane potential, $V$,
of the HH ($A_{\rm syn}=\pm40$ $\mu{\rm A/cm^2}$)
and (d) $V$ of IF1 neurons ($A_{\rm syn}=\pm64$ $\mu{\rm A/cm^2}$);
solid and dashed curves denote
excitatory  and inhibitory cases, respectively,
and scales for $U_{\rm i}$ and $I_{\rm i}$ are arbitrary.
(see text).
 \vspace{0.5cm}

\noindent
{\bf Fig.3}
Responses of the HH neuron to constant ISIs. 
(a) mean ($\mu_{\rm o}$, solid curve), 
RMS ($\sigma_{\rm o}$, dashed curve) 
and the distribution (filled circles) of 
output ISIs  as a function of $A_{\rm syn}$ 
for $\mu_{\rm i}=10$ msec
(the arrow denotes the input ISI);
(b) those as a function of $\mu_{\rm i}$
with $A_{\rm syn}=40 \; \mu{\rm A/cm^2}$
and (c) with $A_{\rm syn}=-40 \; \mu{\rm A/cm^2}$
(dotted lines express $k=1$, 2 and 3).
\vspace{0.5cm}

\noindent
{\bf Fig.4} 
The phase diagrams of the (a) HH, (b) IF0 and (c) 
IF1 neurons in the $\mu_{\rm i}-A_{\rm syn}$ space,
showing the states with integer 
$k \:(=\mu_{\rm o}/\mu_{\rm i})$  
for constant-ISI inputs given by Eqs.(14) and (15),
crosses denoting no outputs.
The results of IF0 and IF1 are shown only for positive
$A_{\rm syn}$ because they cannot fire for inhibitory 
inputs. 
\vspace{0.5cm}

\noindent
{\bf Fig.5} 
The time courses of (a) constant-ISI input and
(b) the membrane potential
of IF1 neuron with $A_{\rm syn}=40$,
(c) $A_{\rm syn}=64$ and
(d) $A_{\rm syn}=120 \; \mu{\rm A/cm^2}$. 
The scale of (a) is arbitrary
and those of (c) and (d)
are shifted by 200 and 400 mV, respectively.
\vspace{0.5cm}

\noindent
{\bf Fig.6}
Responses of the IF0 neuron to constant ISIs. 
(a) mean ($\mu_{\rm o}$, solid curve), 
RMS ($\sigma_{\rm o}$, dashed curve)
and the distribution (filled circles) of 
output ISIs as a function of $A_{\rm syn}$
for $\mu_{\rm i}$=10 msec
(the arrow denotes the input ISI);
(b) those as a function of $\mu_{\rm i}$
(dotted lines express $k=\mu_{\rm o}/\mu_{\rm i}=1$).
\vspace{0.5cm}

\noindent
{\bf Fig.7}
Responses of the IF1 neuron to constant ISIs,
same as in Fig.6. 
\vspace{0.5cm}

\noindent
{\bf Fig.8} 
Histograms of (a) the stochastic input
( $\mu_{\rm i}=10$ msec, $c_{\rm vi}=0.4$),
(b) output ISIs of the HH neuron,
(c) of the IF0 neuron and 
(d) of the IF1 neuron, 
solid (dashed) histograms 
being for $A_{\rm syn}=40 \;(64) \; \mu{\rm A/cm^2}$.
\vspace{0.5cm}

\noindent
{\bf Fig.9}
Responses of the HH neuron to stochastic ISIs; 
(a) $\mu_{\rm o}$ (solid curve), $\sigma_{\rm o}$  (dashed curve)  
and $c_{\rm vo}$ (thin solid curve) as a function of $A_{\rm syn}$
for inputs of $\mu_{\rm i}=10$ msec 
with $c_{\rm vi}=0.4$ and 
(b) with $c_{\rm vi} =1.0$ (the arrow denotes  $\mu_{\rm i}$);
(c) those as a function of $\mu_{\rm i}$
with $A_{\rm syn}=40 \; \mu{\rm A/cm^2}$ 
for inputs of $c_{\rm vi}=0.4$ and 
(d) of $c_{\rm vi} =1.0$;
(e) those as a function of $\mu_{\rm i}$
with $A_{\rm syn}=- 40 \; \mu{\rm A/cm^2}$ 
for inputs of $c_{\rm vi}=0.4$ and 
(f) of $c_{\rm vi} =1.0$.
\vspace{0.5cm}

\noindent
{\bf Fig.10}
Responses of the IF0 neuron to stochastic ISIs;
(a) $\mu_o$ ( solid curve), $\sigma_o$  (dashed curve)  
and $c_{\rm vo}$ (thin solid curve) 
as a function of $A_{\rm syn}$ for inputs of
$\mu_{\rm i}=10$ msec with
$c_{\rm vi}=0.4$ and (b) with $c_{\rm vi}=1.0$ 
(the arrow denoting $\mu_{\rm i}$);
(c) those as a function of $\mu_{\rm i}$ 
with $A_{\rm syn}=64 \; \mu{\rm A/cm}^2$
for inputs of $c_{\rm vi}=0.4$ and 
of $c_{\rm vi}=1.0$.
\vspace{0.5cm}

\noindent
{\bf Fig.11}
Responses of the IF0 neuron to stochastic ISIs, same as Fig.10.
\vspace{0.5cm}

\noindent
{\bf Fig.12} 
$c_{\rm vo}$ against $c_{\rm vi}$ of HH (open squares), IF0 (triangles)
and IF1 neurons (circles) (see text).
\vspace{0.5cm}

\noindent
{\bf Fig.13} 
The $f_{\rm i}-f_{\rm o}$ plot of (a) the HH and (b) IF0 neurons,
dashed curves denoting the extrapolation.
\vspace{0.5cm}


\begin{thebibliography}{99}

\bibitem{Rieke96}F. Rieke, D. Warland, R. Steveninck, and W. Bialek,
{\it Exploring the Neural Code} (MIT press, England, 1996).

\bibitem{Hopfield82}J. J. Hopfield,
Proc. Natl. Acad. Sci. (USA)  {\bf 79} (1982) 2554.

\bibitem{Suga83}N. Suga, W. E. O'Neill, K. Kujirai,
and T. Manabe,
J. Neurophysiol.  {\bf 49} (1983) 1573.

\bibitem{Konishi92}M. Konishi,
Harvey Lect.  {\bf 86} (1992) 47.

\bibitem{Carr86}C. E. Carr, W. Heiligenberg
and G. J. Rose,
J. Neurosci. {\bf 6} (1986) 107.

\bibitem{Eckhorn88}R. Eckhorn, R. Bauer, W. Jordan,
M. Brosch, W. Kruse, M. Munk, and H. J. Reitboeck,
Biol. Cybern. {\bf 60} (1988) 121.

\bibitem{Gray89}C. M. Gray and W. Singer,
Proc. Natl. Acad. Sci. (USA)  {\bf 86} (1989) 1698.

\bibitem{Rolls94}E. T. Rolls and M. J. Tovee,
Proc. Roy. Soc. B {\bf 257} (1994) 9.

\bibitem{Thorpe96}S. Thorpe, D. Fize and C. Marlot,
Nature  {\bf 381} (1996) 520.

\bibitem{Hodgkin52}A. L. Hodgkin and A. F. Huxley,
J. Physiol. {\bf 117} (1952) 500.

\bibitem{Nemoto75}I. Nemoto, S. Miyazaki, M. Saito and
T. Utsunomiya,
Biophys. J. {\bf 15} (1976) 469.

\bibitem{Holden76}A. V. Holden,
Biol. Cybernetics {\bf 21} (1976) 1.

\bibitem{Fohlmeister80}J. F. Fohlmeister, W. J. Adelman, 
and R. E. Poppele,
Biophys. J. {\bf 30} (1980) 79.


\bibitem{Matsumoto80}G. Matsumoto, K. Kim, T. Ueda, and
J. Shimada,
J. Phys. Soc. Jpn. {\bf 49} (1980) 906.

\bibitem{Guttman80}R. Guttman, L. Feldman, and E. Jakobsson,
J. Memb. Biol. {\bf 56} (1980) 9.

\bibitem{Aihara84}K. Aihara, G. Matsumoto, and Y. Ikegaya,
J. Theor. Biol. {\bf 109} (1984) 249.

\bibitem{Matsumoto84}G. Matsumoto, K. Aihara, M. Ichikawa,
and A. Tasaki,
J. Theor. Neurobiol. {\bf 3} (1984) 1.

\bibitem{Arbib95}C. Koch and \"{O}. Bernander, 
{\it The Handbook of Brain Theory and Neural Networks},
ed. M. A. Arbib (MIT press, England, 1995), p129-134.

\bibitem{Albarbanel96}H. D. I. Albarbanel, M. I. Rabinovich,
A. Selverston, M. V. Bazhenov, R. Huerta, M. M. Sushcik,
and L. L. Rubchinskii,  
Physica-Uspekki, 39 {\bf 39} (1996) 337; the table in this
paper provides a good summary of the neuron models.

\bibitem{Maass97}W. Maass,
Neural Comput.  {\bf 9} (1997) 279; 
related references are therein.

\bibitem{Hopfield95}J. J. Hopfield and A. V. M. Herz, 
Proc. Natl. Acad. (USA) {\bf 92} (1995) 6655.

\bibitem{Konig96}P. Konig, A. K. Engel, and 
W. Singer, 
Trends Neurosci.  {\bf 19} (1996) 130.

\bibitem{Abbott97}L. F. Abbott, J. A. Varela, 
K. Sen, and S. B.  Nelson, 
Science  {\bf 275} (1997) 220.


\bibitem{Destexhe97}A. Destexhe,
Neural Comupt. {\bf 9} (1997) 503.

\bibitem{Troyer98}T. W. Troyer, and K. D. Miller,
Neural Computation {\bf 10} (1998) 1047.

\bibitem{Feng97}J. Feng,
Phys. Rev. Lett. {\bf 79} (1997) 4505.


\bibitem{Feng98}J. Feng and D. Brown,
J. Phys. A {\bf 31} (1998) 1239.

\bibitem{Brown99}D. Brown, J. Feng and S. Feerick, 
Phys. Rev. Lett. {\bf 82} (1999) 4731.

\bibitem{Hodgkin48}A. L. Hodgkin,
J. Physiol. {\bf 107} (1948) 165.

\bibitem{Perkel74}D. H. Perkel, B. Mulloney,
Science {\bf 185} (1974) 181.

\bibitem{Hasegawa99}H. Hasegawa, Phys. Rev. E (in press) 
(E-preprint: cond-mat/9906020).

\bibitem{Marsalek97}P. Marsalek, C. Koch and J. Maunsell,
Proc. Natl. Acad. Sci. (USA) {\bf 94} (1997) 735.


\bibitem{Softky92}W. R. Softky and C. Koch,
Neural Comput. {\bf 4} (1992) 643.

\bibitem{Park96}M. Park and S. Kim,
J. Korean Phys. Soc. {\bf 29} (1996) 9.



\bibitem{Horn97}D. Horn and I. Opher,
Neural Comput. {\bf 9} (1997) 1677.


\bibitem{Comment1}
If we adopt $C= \bar{C}= 1.0 \: \mu {\rm F/cm}^2$ as in the HH model
with $\tau_{\rm m}=20$ msec and $V_{\rm t}-V_{\rm r}=20$ mV,
Eq.(1) yields $I_{\rm c}=1.0 \; \mu {\rm A/cm}^2$ which is
much smaller than 6.3 $\mu {\rm A/cm}^2$ of the HH neuron.
On the contrary, if we want to get 
$I_{\rm c} \sim 4 \; \mu{\rm A/cm}^2$ with 
$C = 1.0 \: \mu {\rm F/cm}^2$ 
and $\tau_{\rm m}=20$ msec, 
we have to choose $V_{\rm t}-V_{\rm r}=80$ mV
which is too large compared with the conventional value of 20 mV.
The parameters chosen in the text are the compromised ones.


\bibitem{Tsodyks98}M. Tsodyks, K. Pawelzik adn H. Markram,
Neural Compt. {\bf 10} (1998) 821.

\bibitem{Shadlen94}M. Shadlen and W. T. Newsome,
Curr. Opin. Neurobiol. {\bf 4} (1994) 569.




\bibitem{Usher94}M. Usher, M. Stemmer, C. Koch and Z. Olami,
Neural Comp. {\bf 6} (1994) 795.

\bibitem{Softky95}W. R. Softky,
Curr. Opin. Neurobiol. {\bf 5} (1995) 239.

\bibitem{Troyer97}T. W. Troyer and K. D. Miller,
Neural Computation {\bf 9} (1997) 971.


\bibitem{Softky93}W. R. Softky and C. Koch,
J. Neurosci. {\bf 13} (1993) 334.





\bibitem{Gutkin98}B. S. Gutkin, and G. B. Ermentrout,
Neural Computation {\bf 10} (1998) 1047.


\bibitem{McCormick85}D. A. McCormick, B. W. Connoers,
J. W. Lighthall, and D. A. Price,
J. Neurophysil. {\bf 54} (1985) 782.



\bibitem{Hippo}J. Rinzel, D. Terman, X. Wang,
and B. Ermentrout,
Science  {\bf 279} (1998) 1351; related references 
therein.
















































 





\end{thebibliography}
\end{document}